\documentclass[numbers]{elsarticle}
\usepackage{latexsym}
\usepackage{natbib}
\usepackage[latin2]{inputenc}
\usepackage[hidelinks]{hyperref}
\usepackage{longtable, lineno}

\makeatletter
\def\ps@pprintTitle{%
	\let\@oddhead\@empty
	\let\@evenhead\@empty
	\def\@oddfoot{\centerline{\thepage}}%
	\let\@evenfoot\@oddfoot}
\makeatother

\begin{document}

\title{High-Resolution Directed Human Connectomes and the Consensus Connectome Dynamics}


\author[p]{Balázs Szalkai}
\ead{szalkai@pitgroup.org}
\author[p,s]{Csaba Kerepesi}
\ead{kerepesi@pitgroup.org}
\author[p]{Bálint Varga}
\ead{balorkany@pitgroup.org}
\author[p,u]{Vince Grolmusz\corref{cor1}}
\ead{grolmusz@pitgroup.org}
\cortext[cor1]{Corresponding author}
\address[p]{PIT Bioinformatics Group, Eötvös University, H-1117 Budapest, Hungary}
\address[u]{Uratim Ltd., H-1118 Budapest, Hungary}
\address[s]{Institute of Computer Sceince and Control of the Hungarian Academy of Sciences}

\date{}


\begin{abstract}
Here we show a method of directing the edges of the connectomes, prepared from diffusion tensor imaging (DTI) datasets from the human brain. Before the present work, no high-definition directed braingraphs (or connectomes) were published, because the tractography methods in use are not capable of assigning directions to the neural tracts discovered. Previous work on the functional connectomes applied low-resolution functional MRI-detected statistical causality for the assignment of directions of connectomes of typically several dozens of vertices. Our method is based on the phenomenon of the ``Consensus Connectome Dynamics'' (CCD), described earlier by our research group. In this contribution, we apply the method to the 423 braingraphs, each with 1015 vertices, computed from the public release of the Human Connectome Project, and we also made the directed connectomes publicly available at the site \url{http://braingraph.org}. We also show the robustness of our edge directing method in four independently chosen connectome datasets: we have found that 86\% of the edges, which were present in all four datasets, get the very same directions in all datasets; therefore the direction method is robust, it does not depend on the particular choice of the dataset.
We think that our present contribution opens up new possibilities in the analysis of the high-definition human connectome: from now on we can work with a robust assignment of directions of the connections of the human brain.
\end{abstract}

\maketitle

\section*{Introduction} 

Diffusion tensor imaging (DTI) datasets are widely used today for the construction of connectomes or braingraphs. The nodes (or vertices) of these graphs correspond to anatomically identified \cite{Desikan2006,Fischl2012}, small (1-1.5 cm$^2$) areas of the gray matter, and two nodes are connected by an undirected edge if neural fiber tracts are discovered, connecting the gray matter areas, corresponding to the two nodes \cite{Hagmann2008,Szalkai2015}.

The considerable advantage of the graph approach to the analysis of the human brain is that the graphs retain the relevant anatomical connection information from the MRI data, but they do not contain the --- mostly irrelevant --- data on the exact orbits of the neural tracts in the white matter.  

One of the most frequently used and largest diffusion MRI datasets are the public releases of the Human Connectome Project (HCP) \cite{McNab2013}. Numerous publications are applying HCP data in different contexts, e.g., \cite{delaIglesia-Vaya2011, Szalkai2016b, Glasser2013, Szalkai2016c, Herrick2014, Smith2013b, Szalkai2016a}.

\subsection*{Directing the Connectome: fMRI-Based Methods}

Diffusion tensor imaging methods cannot assign directions to the fiber tracts, and, consequently, to the edges of the connectomes mapped. Tractography methods can discover the orbits, that is, the geometric curves of the fiber tracts in the white matter, but --- presently --- cannot determine the correct one from the two possible directions of any given fiber tract. Consequently, the connectomes of braingraphs that are prepared from the diffusion MR imaging data are {\em undirected graphs}.

Functional MRI, on the other hand, is capable of mapping the temporal sequences of the activity of larger brain areas. Therefore, it seems to be possible to assign directions to these functional or effective connections between those larger areas of the gray matter.  Several authors have attempted assigning directions to the edges of the connectomes using temporal sequences of activity changes, mostly on resting state fMRI or electro-physiology data, by applying a broad spectrum of statistical causality detection methods and models (e.g., \cite{Roebroeck2005,Penny2004,Kim2011,Ching2013,Timme2014, Schelter2006,Ren2010,Stetter2012,Gerhard2013,Zaytsev2015,Smith2011,Pillow2011}). The general approach of the articles listed can be described as follows: the temporal sequence of cerebral activity changes are translated to causality relations by statistical methods, and these causality relations are used to direct the connections between brain regions. 

Since both the fMRI data and the non-invasive electro-physiological measurements have poor spatial resolution, usually the goal of these methods is not directing the edges of the diffusion tensor imaging-based, high-definition anatomical connectomes, but rather the functional connectomes on several dozens of vertices, corresponding to large areas of the gray matter, where haemodynamic activity changes were observed.

By the best of our knowledge, no significant attempts were made to direct the edges of structural connectomes (i.e., connectomes measured by DTI) with hundreds of vertices by these methods. One possible reason is due to the poor resolution of the fMRI: it is impossible to decide that if a causality is very probable between two cortical areas $A$ and $B$ in the direction from $A$ to $B$, then which of the hundreds of the edges, running between $A$ and $B$ should be directed from $A$ to $B$: theoretically, even one such connection may trigger the observed haemodynamic activity in $B$.

\subsection*{Consensus Connectomes}

It is unusual in graph theory that hundreds of different graphs on the very same set of vertices are encountered. When we are working on braingraphs constructed from diffusion tensor imaging (DTI) data, this is the usual scenario: the cerebral areas of the different human subjects can be corresponded to the same reference brain map \cite{Fischl2012,Desikan2006}, and, consequently, the graph edges in the connectome can be compared between subjects, since the vertices or nodes are named in each subject according to the very same anatomical brain map. This process is, in fact, a non-trivial refinement of the assignment of anatomical nomenclature to cerebral areas in printed brain anatomy atlases that appeared in the last several hundred years (e.g., \cite{ValverdedeAmusco1560, Laskowski1894, Kiss1987}).

Consensus Connectome Dynamics (CCD) is a surprising phenomenon that was observed after our construction of the Budapest Reference Connectome Server  \cite{Szalkai2015a,Szalkai2016,Kerepesi2015b}. For the description of the CCD phenomenon, we need to clarify some functions of the Budapest Reference Connectome Server.

The webserver at the address \url{http://connectome.pitgroup.org} is capable of finding and visualizing connectome edges that are present in more than a pre-defined number of connectomes. More exactly, let $n$ denote the number of all connectomes processed by the server ($n=6$ in version 1.0, $n=96$ in version 2.0 and $n=418$ or $n=477$ --- depending on other settings --- in version 3.0).  For any $k$: $0\leq k \leq n$, we say that {\em the frequency of an edge} is $k$ if the edge is present in exactly $k$ connectomes out of the $n$ ones \cite{Kerepesi2016}. Clearly, if we have $n$ connectomes then the frequency of any edge is between $0$ and $n$: it is $0$ if the edge does not appear in any connectome at all; and it is $n$ if it appears in all the $n$ connectomes. The $k${\em-consensus connectome} contains all the edges with a frequency greater than or equal to $k$: that is, each edge in the $k${\em-consensus connectome} is present in at least $k$ individual connectomes.

The Budapest Reference Connectome Server is capable of generating $k${\em-consensus connectomes} in CSV and GraphML formats for further processing and independent visualization, and on the website users can also quickly view the resulting consensus connectomes. 

\subsection*{Consensus Connectome Dynamics}

After the publication of the article describing the Budapest Reference Connectome Server \cite{Szalkai2015a}, a surprising property of the server was recognized \cite{Kerepesi2015b}: when we generate $k$-consensus connectomes for $k=n$, next for $k=n-1$,$k=n-2$,..., and last for $k=1$, then, naturally, more and more edges appear in the graphs (since the inclusion condition is weakened in every step). The surprising phenomenon is that the new edges do not appear randomly in the graph, but new edges are mostly connected to the already existing edges, forming a growing, tree-like structure as it is shown on the video \url{https://youtu.be/yxlyudPaVUE}, \cite{Kerepesi2015b}. This ``dynamical'' observation is quantified on Fig. 2 of \cite{Kerepesi2015b}, and it is statically visualized on a very large component-tree at the address \url{http://pitgroup.org/static/graphmlviewer/index.html?src=connectome_dynamics_component_tree.graphml} (for the detailed --- and not entirely obvious ---  description of the component-tree we refer to \cite{Kerepesi2015b}).

In \cite{Kerepesi2015b} we called this phenomenon ``Consensus Connectome Dynamics'', and hypothesized that the particular order of the appearance of the graph edges describes the order of growth of the axonal fibers of these connections in the individual brain development. 

If the newly formed edges - i.e., fibers of axons - almost always are connected to the already developed edges - i.e., fibers of axons - then this hypothesis satisfyingly explains the Consensus Connectome Dynamics phenomenon, visualized at \url{https://youtu.be/yxlyudPaVUE}: 
\begin{itemize}
	\item Connections, represented by the graph edges that are present in most connectomes were grown first (e.g., edges in the $n$- or $n-1$-consensus connectomes). 
	\item Next, the newer and newer edges were grown to be connected to the vertices of the older edges of the growing, tree-like structure, forming the $k$-consensus connectomes, for decreasing $k$ values. 
\end{itemize}

The newer and newer edges have gradually larger deviation, since small differences in the edge frequencies are added up in the growing structure: Clearly, if an edge $e$ is not present in 20\% of the graphs, then for another edge, say $f$, which connects to $e$, will hold that the $e,f$ edge-pair cannot be present in at least 20\% of the graphs, and most probably, the $e,f$ pair will be missing from much more than 20\% of all graphs, since $f$ will be missing in some graphs, containing $e$.  Consequently, the frequencies of the newer edges will be less than the frequencies of older edges. Therefore, we believe that the visualization at \url{https://youtu.be/yxlyudPaVUE}, describes {\em not only} the edges with gradually smaller frequencies, but also the order of growth of the connections in the human brain.

\section*{Results and Discussion}

In the present work, we are assigning directions to some of the edges of braingraphs, depending on the appearance of that edge in the CCD. 

The edge-directing method was first described in \cite{Kerepesi2015b}; here we clarify the method, and make the 423 directed-edge connectomes publicly available at the site \url{http://braingraph.org}.

We have two requirements on the edge-directing method, based on Consensus Connectome Dynamics:

- {\bf Feasibility:} While our direct knowledge is very limited on the axonal development of the human brain in general and the directions of fiber tracts in the connectome in particular (\cite{DeCarlos1992,Lewis2013}), we need to build onto the available biological observations in developing our method;

- {\bf Robustness:} The directions, assigned to the edges of the connectomes need to be as independent as possible from the selected, specific sets of connectomes in the connectome sets in CCD trials. In other words, the CCD phenomenon needs the data of dozens or hundreds of connectomes; based on these connectomes and the CCD phenomenon, we assign directions to some of the edges of the individual connectomes. However, we want these directions assigned to the edges of the individual braingraphs as independent as possible from the {\em particular} choice of the several dozen or several hundred connectomes included in the CCD probe.

Here we suggest an (a) Feasible and (b) Robust edge-directing method.

\subsection*{Feasibility}

A large part of the axonal connections in the brain of mammals is developed in the embryonic state or the first several weeks after birth \cite{Lewis2013}. Microscopic studies of developing  rat brains show that a considerable number of axons of the cortex (from layer V cortical neurons), in the early brain development, grow in the direction of subcortical regions \cite{DeCarlos1992}, strengthening our CCD observation that neurons of the cortex are connected later to the subcortical structures than the connections develop between those subcortical structures. 

Another important observation that forms a base of our method of directing the axonal fibers is the retrograde signaling of post-synaptic activation that stabilizes the newly formed synapses, and, consequently, prevents the retraction of the newly formed axon branches: work by numerous groups (e.g., \cite{Hu2005, Cline1989, Rajan1999, Schmidt2000, Ruthazer2004,Zou1996a,Schmidt2004}) witness this phenomenon.

Based on these observations, we make the following hypothesis: Neurons that are not connected to any other neuron yet, grow axons to reach other neurons. 
\begin{itemize}
	\item[(i)] If the branch of the axon reaches a neuron that is not connected to other neurons yet, it retracts, since the post-synaptic activity of the un-networked neuron does not stabilize the connection (i.e., the synapse).
	
	\item[(ii)] If the branch of the axon reaches an already ``networked'' neuron, that is, a neuron with connections to other neurons, then the post-synaptic activity of the ``networked'' neuron stabilizes the synapse and, consequently, the new connection.  
\end{itemize}

Since the neuronal-level human braingraph with 80 billion neurons as vertices is unavailable today, we re-phrase the hypothesis above for the ROIs that form the 1015 vertices of our braingraphs, as follows:

ROIs that are not connected to any other ROI yet, grow axonal fibers to reach other ROIs as observed in the CCD phenomenon. 
\begin{itemize}
	\item[(iii)] If the fiber reaches an ROI that is not connected to other ROIs yet, the fiber retracts, since the post-synaptic activities of the un-networked neurons do not stabilize the neuronal connections.
	
	\item[(iv)] If the fiber reaches an already ``networked'' ROI, that is, an ROI with connections to other ROIs, then the post-synaptic activities of the ``networked'' neurons stabilize the synapses and the new connection will survive.  
\end{itemize}

We stress that -- instead of the mainly unknown order or temporal sequence of axonal development on the system level -- we consider the phenomenon of the Consensus Connectome Dynamics, and we direct the edges of the connectomes according to the hypotheses (iii) and (iv) above.

\subsection*{Robustness}

The robustness of the edge direction method is built into the edge direction algorithm, as it is detailed in the ``Methods'' section.

\section*{Methods}

The braingraphs in this study were constructed from the data of the Human Connectome Project \cite{McNab2013}, using the workflow described in \cite{Szalkai2015a}. The undirected braingraphs have 1015 vertices each; the whole set of 423 braingraphs can be downloaded from the site \url{http://braingraph.org/download-pit-group-connectomes/}.

For verifying the robustness of the edge direction method, we first divided the braingraphs of the 423 subjects into four groups, the first contained 105, the others 106 graphs each. The graphs of each four groups were used -- separately -- to reproduce the CCD phenomenon, and based on the order of the appearance of the edges in the four different CCD probes, we assigned directions to some of the edges. Had the resulting directed graphs been radically different for the four groups, it would have indicated that our method is not a robust method of directing the edges of braingraphs. 

As we detail below, it turned out that the four directed braingraphs were very similar to each other. After obtaining these four directed graphs, we used a majority-like approach to direct the edges of the consensus brain graph for the whole population, and, in a next step, the edges of the individual braingraphs.

\subsection*{Computing Consensus Braingraphs:} For each group, for $i= 1, 2,3,4$, we defined the consensus braingraph $G_i$ as the multi-union of the individual connectomes. That is, for each possible graph edge we counted the subjects whose connectomes contained the specific edge. This number ranges from $0$ to $N_i$, where $N_i=106$, is the number of subjects in the given group. We defined the consensus braingraph as a simple graph containing all edges that were present in at least one of the subjects, where the edges are labeled (weighted) with the number of individual connectomes containing them.

\subsection*{Directing the edges with BFS:} We defined edge directions on this brain graph $G_i$ as follows. For each $k$ from $N_i$ down to $1$, we defined $G_{ik}$ as the subgraph of $G_i$ containing all edges with frequency at least $k$. This means all the edges which were present in at least $k$ subjects within the $i$th group. Suppose that $G_{ik}$ is already processed, and we want to process the graph $G_{i(k-1)}$, which is an augmentation of $G_{ik}$ with precisely those edges which are present in exactly $k-1$ individual graphs. Since $G_{ik}$ is already processed, our task is now to direct these newly added edges only. To achieve this, we launch a BFS (breadth-first search) in $G_{i(k-1)}$, with $G_{ik}$ as the source. This means that, for each node in $G_{i(k-1)}$, we calculate the distance of this node from $G_{ik}$ within the graph $G_{i(k-1)}$. This distance is at least $0$, is precisely $0$ for the nodes of $G_{ik}$, is finite for the nodes connected to $G_{ik}$ with a path of edges, and is infinite for the nodes of $G_{i(k-1)}$ not connected to $G_{ik}$. Nodes with infinite distance do exist because sometimes isolated edges or new, small graph components appear in $G_{i(k-1)}$.

After calculating the node distances, each edge in $G_{i(k-1)}$ will connect either two nodes of the same distance, or two nodes with a distance differing by exactly 1. If an edge connects $u$ with $v$, and the distance of $v$ is exactly 1 less than the distance of $u$, then we direct that edge from $u$ to $v$. If an edge connects two equidistant nodes, then we leave this edge undirected and it will not acquire a direction in this consensus brain graph, not even in the subsequent steps.

This way, when we reach frequency $k=1$ and have processed $G_{i1} = G_i$, we obtain a {\em partially directed} version of $G_i$. Some of its edges will become directed, and the remaining will be left undirected. As we do this for all four of the groups, we obtain four directed brain graphs on the same set of nodes.

\subsection*{Directed Consensus Connectome:} We unify these directed connectomes into one big directed consensus connectome as follows. If an edge is directed in the same direction in at least 2 of the 4 consensus graphs, and is undirected or directed in the other direction in at most 1 of the graphs, then the direction of this edge is defined as the most common value. For example, if an edge is directed forward in 2 graphs, backward in 1 graph, and is absent from the remaining graph, then the edge will become forward-directed in the result graph. On the other hand, if that edge is directed forward in 2 graphs, backward in 1 graph, and is undirected in the remaining graph, then its direction is considered ambiguous, and this edge will remain undirected in the result graph.

By running this algorithm, we could direct 48324 of the 71783 edges of the consensus brain graph for the whole population. This means that we could assign a direction to 67\% of all the edges which were present in at least one of the braingraphs. We think that 67\% as a very significant portion of the edges, since the edges include even those sporadic connections which were present in only one brain graph and thus cannot be oriented reliably. If we count only those edges which were present in all 4 consensus connectomes (there were 31873 such edges), then we see that 26305 of them were directed the same way in all 4 groups. This means that 82\% of these 31873 edges acquired the same orientation in all 4 groups.

This proves the robustness of our method. Additionally, these numbers also show that the CCD phenomenon is robust, too. 

\subsection*{How can we direct the individual braingraphs?} In the previous subsection we have presented a CCD-based method to direct most of the edges of the consensus connectome in a robust way. The edges of the individual braingraphs can be directed by applying the directions of the edges of the (partially) directed consensus braingraph as follows: 

Let us consider an individual braingraph $G$ and the directed consensus braingraph $\bar G$. If an edge $e$ of $G$ is directed in $\bar G$, then let us direct $e$ in the same way in $G$, too. If edge $e$ of $G$ is undirected in $\bar G$, then $e$ will remain to be undirected in $G$, too.

\section*{Data availability:} The Human Connectome Project's MRI data that were applied in the present work is available at: 
\url{http://www.humanconnectome.org/documentation/S500} \cite{McNab2013}. 

\noindent The graphs (both undirected and directed) that were prepared by us from the HCP data can be downloaded at the site \url{http://braingraph.org/download-pit-group-connectomes/}. 

The Budapest Reference Connectome Server \cite{Szalkai2015a,Szalkai2016} is available at \url{http://connectome.pitgroup.org}. On that site, the reader can independently verify the phenomenon of the Consensus Connectome Dynamics by (i) choosing ``Show options'' (ii) and moving from right to left the ``Minimum edge confidence'' slider (iii) and observing the buildup of the edges on the visualization panel. If graphics problems appear then it is suggested to use another browser, e.g., Chrome. 

The Consensus Connectome Dynamics is visualized on a video animation on \url{https://youtu.be/yxlyudPaVUE} for the whole brain and on    \url{https://youtu.be/wBciB2eW6\_8} for the frontal lobe only.

The individual braingraphs with directed edges can be accessed at the site \url{http://braingraph.org/download-pit-group-connectomes/}. Each graph refers to the original HCP ID in its filename.

\section*{Acknowledgments}
Data were provided in part by the Human Connectome Project, WU-Minn Consortium (Principal Investigators: David Van Essen and Kamil Ugurbil; 1U54MH091657) funded by the 16 NIH Institutes and Centers that support the NIH Blueprint for Neuroscience Research; and by the McDonnell Center for Systems Neuroscience at Washington University.

\section*{References}


\end{document}